\documentclass[a4paper]{article}
\usepackage{graphicx}
\usepackage{lmodern}
\usepackage[latin1]{inputenc}
\usepackage[affil-it]{authblk}
\usepackage{calc}
\usepackage{amsmath}
\usepackage{amssymb}
\usepackage{relsize}
\usepackage{multirow}
\usepackage{rotating}
\usepackage{bm}
\usepackage{url}
\usepackage{multicol}
\usepackage{array}
\usepackage{booktabs}
\newcommand{\head}[1]{\textnormal{\textbf{#1}}}
\newcommand{\normal}[1]{\multicolumn{1}{l}{#1}}

\makeatletter
\def\@maketitle{%
  \newpage
  \null
  \vskip 2em%
  \begin{center}%
  \let \footnote \thanks
    {\Large\bfseries \@title \par}%
    \vskip 1.5em%
    {\normalsize
      \lineskip .5em%
      \begin{tabular}[t]{c}%
        \@author
      \end{tabular}\par}%
    \vskip 1em%
    {\normalsize \@date}%
  \end{center}%
  \par
  \vskip 1.5em}
\makeatother

\title{Identified Two-particle Correlations and Quantum Number Conservations in p-p and Pb-Pb Collisions at LHC Energies}
\author[1]{Gyula Benc\'edi\thanks{\texttt{bencedi.gyula@wigner.mta.hu}; Corresponding author}}
\author[1]{Gergely G\'abor Barnaf\"oldi\thanks{\texttt{barnafoldi.gergely@wigner.mta.hu}}}
\author[2]{Levente Moln\'ar\thanks{\texttt{Levente.Molnar@cern.ch}}}
\affil[1]{Wigner Research Centre for Physics of the Hungarian Academy of Sciences, Budapest, Hungary}
\affil[2]{Institut Pluridisciplinaire Hubert Curien, Strasbourg, France}

\date{Dated: \today}

\begin{document}

\maketitle

\begin{abstract}
In this paper we continue the investigation of the effect of quantum number conservations of pions, kaons, and protons, with very high transverse momenta (up to 25 GeV/c), during parton fragmentation and hadronization in p-p and Pb-Pb collisions at LHC energies~\cite{Acconcia:2013ptg,Bencedi:2014xka}. The strength of the conservation effects are studied by identified two-particle correlations in Monte Carlo generated events in the mid-rapidity region ($|\eta| < 1$). The simulated p-p events were generated with PYTHIA 8~\cite{Sjostrand:2007gs}, using its main default settings, at $\sqrt{s}=200$~GeV, $\sqrt{s}=2.76$~TeV, $\sqrt{s}=7$~TeV, and $\sqrt{s}=14$~TeV. In parallel to this, HIJING 1.36~\cite{GyulassyHijing} was used to generate Pb-Pb events at $\sqrt{s_{\rm NN}}=2.76$~TeV with centralities $0-10\%$, $30-40\%$ and $80-90\%$. The extracted identified associated hadron spectra for charged pion, kaon, and proton show identified trigger-hadron dependent splitting between oppositely charged associated particle species in any nucleus-nucleus collisions. The Pb-Pb data exhibits a peculiar splitting pattern as a function of the transverse momentum of the associated particle $p_{T,assoc}$ both on the near and away side that is different from the patterns observed in p-p collisions. The splitting shows smooth evolution with collision energy and event multiplicity in p-p collisions while in Pb-Pb collisions different trend were observed for kaons and protons.
\end{abstract}

\section{Introduction}
\label{sec:1}

Two-particle angular correlation is an efficient and powerful tool to study the properties of the underlying physics processes of particle production in different collision systems, like proton-proton or heavy-ion collisions. Using identified particles we can check those kind of conservation laws which are related to the different quark and flavour content of the processes~\cite{Bencedi:2014xka}. This can help us to understand the specific hadronization processes in more details. Furthermore, focusing on the correlations of identified particles one can study recombination effects~\cite{Fries:2003kq, Hwa:2003ic} besides the parton fragmentation in the hadronization mechanisms. We used in our analysis the most known Monte Carlo generators (PYTHIA~\cite{Sjostrand:2007gs} and HIJING~\cite{GyulassyHijing}) of the high energy physics community. This can be practically helpful to verify the validity and the prediction power of the studied physical quantities of this study to experimental data.

In this paper we continue our earlier work~\cite{Bencedi:2014xka} on the investigation of the particle production and the quantum number conservation of identified hadron-triggered angular correlation of identified hadrons at high transverse momenta, up to $p_{T}=25$~GeV/c. This time the strength of the conservation effects is studied as a function of collision energy and event multiplicity in p-p collisions and as a function of centrality in Pb-Pb collisions.

This proceeding is organized as follows: Sec.~\ref{sec:2} briefly presents the Monte Carlo data samples and analysis settings which were used in this study. In Sec.~\ref{sec:3} we show the obtained results for relative yields of unlike-sign associated particle pairs in specific physical configurations in p-p and Pb-Pb collisions. Finally, Sec.~\ref{sec:4} summarizes the results of this proceeding and gives outlook for possible further analysis of the subject.

\section{Monte Carlo simulations of identified \\two-particle azimuthal correlations in \\p-p and Pb-Pb collisions}
\label{sec:2}

\subsection{Monte Carlo data sets and analysis cuts}
\label{subsec:0}

As in our earlier study, here we also consider generated identified particles: $\pi^{\pm}$, $K^{\pm}$, $p$, $\bar{p}$, and the charged hadrons $(h^\pm)$ as a reference~\cite{Bencedi:2014xka}.

Event generators PYTHIA~8~\cite{Sjostrand:2007gs} and HIJING~1.36~\cite{GyulassyHijing} were used for the production of p-p and Pb-Pb event samples. PYTHIA~8 was used with its default settings (tune 4C), while HIJING event generation included the simulated effects of quenching and shadowing. The detailed data sets for the generation of p-p and Pb-Pb samples can be found in Table~\ref{tab:tab1}.
All charged final-state particles were kept for analysis from the Monte Carlo events, nevertheless the trigger particles and associated particles used for the two-particle correlations measurements were limited in pseudorapidity in order to take into account the sensitive geometry of today's experimental detectors. Trigger particles with $|\eta_{trig}|>0.5$ and the associated particles with $|\eta_{assoc}|>1$ were rejected.
In order to exclude the low-momentum region from the bulk effects, the transverse momenta of the trigger particles were considered above $p_{T,trig}>2$~GeV/c in all the cases. The associated particle momenta were chosen such that $p_{T,assoc} < p_{T,trig}$ to avoid double counting. The uncorrelated background was subtracted by applying the ZYAM method~\cite{zyam}.

\vspace{1mm}

\begin{table}[!htb]
\begin{center}
\small
\begin{tabular}{c c r c r}
\toprule
\textbf{Collision system} & & $\sqrt{s_{NN}}$ & & \textbf{Statistics} \\
\midrule
p-p 		& & $14$~TeV	& & 500~M\\
p-p		& & $7$~TeV	& & 100~M\\
p-p		& & $2.76$~TeV	& & 100~M\\
p-p		& & $200$~GeV	& & 500~M\\
\midrule
Pb-Pb~(0-10\%)	& & $2.76$~TeV    & & 4~M\\
Pb-Pb~(30-40\%)	& & $2.76$~TeV    & & 2~M\\
Pb-Pb~(80-90\%)	& & $2.76$~TeV    & & 10~M\\
\bottomrule
\end{tabular}
\end{center}
\caption{The list of Monte Carlo generated data sets used for our analysis. Proton-proton events were generated with different collision energies, while Pb-Pb events simulated with different centralities as indicated in the parenthesis. See the text for further details.}
\label{tab:tab1}
\end{table}

\normalsize

\subsection{The definition of the associated per trigger yield}
\label{subsec:0b}

Measuring quantum number conservation requires constraints both on the trigger and the associated particle directions to tighten the sensitivity of the measurement. The associated per trigger yield is defined with the following quantity as a function of pseudorapidity difference $\Delta\eta= \eta_{trig}-\eta_{assoc}$ and azimuthal angle difference $\Delta\phi = \phi_{trig}-\phi_{assoc}$:
\begin{equation}
\label{eq:assyield}
\frac{\mathrm{d^{2}}N}{\mathrm{d}(\Delta\eta)\mathrm{d}(\Delta\phi)}(\Delta\phi,\Delta\eta,p_{T,trig},p_{T,assoc}) = \frac{1}{N_{\mathrm{trig}}}\cdot\frac{\mathrm{d}N_{\mathrm{assoc}}}{\mathrm{d}(\Delta\eta)\mathrm{d}(\Delta\phi)}~,
\end{equation}
where $N_{\mathrm{trig}}$ and $N_{\mathrm{assoc}}$ is the number of trigger and associated particles per event, respectively. This quantity was measured in several $p_{T,trig}$  and $p_{T,assoc}$ intervals. The azimuthal angle correlations in this study are always projected within the pseudorapidity difference $|\Delta\eta|<1$. The associated per trigger yield on the near side was extracted from the $\Delta\phi$ interval $|\Delta\phi|<\pi/2$ and the away side from the $\Delta\phi$ interval $\pi/2 < \Delta\phi < 3\pi/2$.

\subsection{The PID-associated spectra}
\label{subsec:1}

To test the quantum number conservation the identified associated spectra of the associated particles (hereafter called PID-associated spectra) have been plotted. We have investigated the PID-associated raw spectra using $\pi^{\pm}$, $K^{\pm}$, $p$, $\bar{p}$, and charged hadron $h^{\pm}$ triggers.

	\begin{figure}[!htb]
		\centering
		{\includegraphics[bb = 0 0 473 237, width=0.75\linewidth]{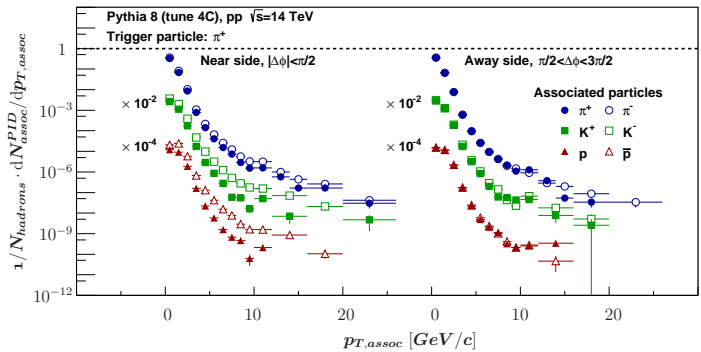}\label{fig:1}}
	    	\caption{(Color online) $\pi^{+}$-triggered associated particle spectra for near- and away side particles scaled by integral of the charged hadron identified associated spectra in simulated p-p events at $\sqrt{s}=14$~TeV. Additional scaling was applied to the spectra for better visibility. See the text for the details.
		}
		\label{fig:PID-triggered_spectra}
	\end{figure}

Figure~\ref{fig:PID-triggered_spectra} shows the identified particle yields $\mathrm{d} N^{PID}_{assoc}/ \mathrm{d} p_{T,assoc}$ for the identified associated particles up to $p_{T,assoc}<26$ GeV/c in p-p collisions at $\sqrt{s}=14$~TeV. The kaon (green symbols) and proton (red symbols) spectra are scaled for better visibility.
The PID-associated spectra have been plotted for positively charged pions, as trigger particles, selected from the transverse momentum range \mbox{$2$ GeV/c $ < p_{T,trig} < 25$ GeV/c}.

On the left side of Figure~\ref{fig:PID-triggered_spectra} one can see the $p_{T,assoc}$ spectra for pions, kaons and protons for the 'near side' case. Right side shows the spectra for the 'away side' case. Full symbols refer to particles with positive charge and negatively charged particles are indicated with open symbols.

The PID-associated spectra of the different associated particle species show the expected evolution with the transverse momentum. The yields of the PID-associated spectra significantly decrease with the selection of charged pion, kaon and proton triggers.

\vspace{2mm}

\begin{table}[!htb]
\centering
\smaller
\begin{tabular}{@{}l*2{>{}c}%
  c<{}@{}}
  \toprule[1pt]
  & \multicolumn{1}{c}{\head{Trigger~/~Assoc}} &
    \multicolumn{2}{c}{\head{Strength of effect}}\\
  & \normal{} & \normal{\head{Near side}} &
  \normal{\head{Away side}} \\
  \cmidrule(lr){2-2}\cmidrule(l){3-4}
  \multirow{3}{*}{}
  & $\pi^{+}$~/~($\pi^{+},\pi^{-}$)	&  $\sim2$ & $\sim1$\\
  & $K^{+}$~/~($K^{+},K^{-}$)		&  $\sim5$ & $\sim1$\\
  & $p$~/~($p,\bar{p}$)			&  $\sim10$ & $\sim1$\\[1ex]
  \bottomrule[1pt]
\end{tabular}
\caption{Qualitative yield differences between positive and negative associated particles on the near- and away sides, listed here on Fig.~\ref{fig:PID-triggered_spectra}.}
\label{tab:tab2}
\end{table}

On a qualitative basis one can see on the left side of Fig.~\ref{fig:PID-triggered_spectra} the yield differences between the positive and negative associated particles. Looking at the right side of the figure, no clear difference can be observed between associated particles on the away side. We listed these qualitative observations in Table~\ref{tab:tab2}. It shows essentially that the largest yield difference we observe is for protons and the smallest one is present for pions, as open and full symbols are splitting towards higher $p_{T}$ values.\\

We note here, that similar plots can also be obtained using other type of trigger hadron species, as we do in the next subsection~\ref{subsec:2}.

\subsection{Identified particle ratios}
\label{subsec:2}

This part was already discussed in~Ref.~\cite{Bencedi:2014xka} in detail, thus here we remind the reader of the main definition and features of the defined quantity.\\
The charge ($C$) and flavour specific or other quantum numbers, like baryon number ($B$) and strangeness ($S$), of generated particles are expected to hold during the hadronization and its traces can be seen on the final PID-associated spectra. In order to quantify and magnify the expected quantum number conservation effects, we define the following ratio:
\begin{equation}\label{eq:21}
\frac{\mathrm{d}N^{PID}_{assoc}}{\mathrm{d}p_{T,assoc}}~\Big/~\frac{\mathrm{d}N^{hadron}_{assoc}}{\mathrm{d}p_{T,assoc}}~.
\end{equation}
This quantity can be understood as follows: in the nominator we consider the case when we are interested in the yield of the identified associated particles having an identified particle as trigger. While in the denominator we calculate the yield of identified associated particles without any PID information on the trigger particles, i.e. using charged hadrons as triggers.

	\begin{figure}[!h]
		\centering
		\includegraphics[width=0.75\textwidth]{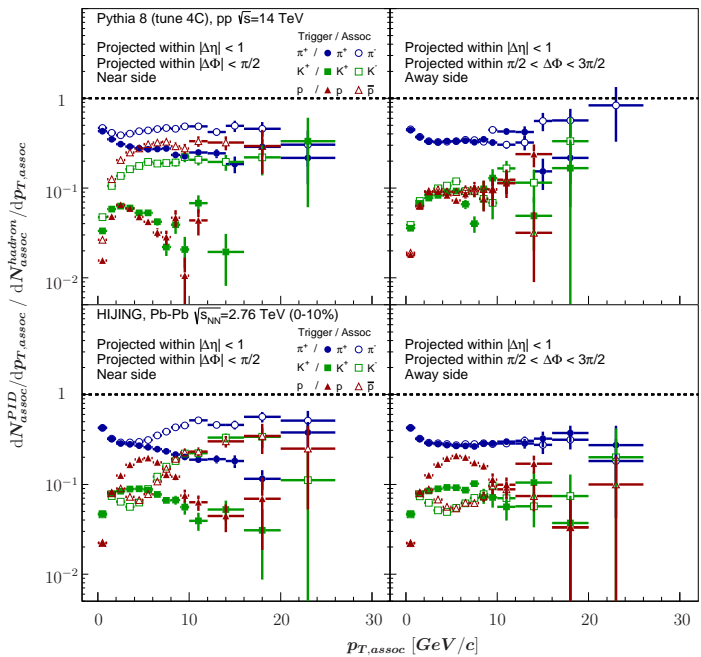}
		\caption{(Color online) PID-triggered associated particle yields relative to the charged hadron-triggered associated yields. The upper row shows the PYTHIA simulation in p-p and the lower row shows the $0-10\%$ central Pb-Pb HIJING events. The \textit{left} column corresponds to the near side ($\pi/2 < |\Delta\phi|$) correlations and the \textit{right} column corresponds to the away side ($\pi/2 < \Delta\phi < 3\pi/2$) correlations. See the text for details.}
\label{fig:PID-triggered_ratio}
	\end{figure}

In Figure~\ref{fig:PID-triggered_ratio} the ratios, defined by Eq.~(\ref{eq:21}), for six identified trigger~/~identified associated particle species are shown: $\pi^{+}/\pi^{+}$, $\pi^{+}/\pi^{-}$, $K^{+}/K^{+}$, $K^{+}/K^{-}$, $p/p$, and $p/\bar{p}$. We used only the like-sign ($+/-$), unlike-sign ($+/+$) particle combinations, since we are interested in the flavour balance between the associated particles regarding specific quantum number.\\
The upper panel shows the ratio of the yields defined by Eq.~(\ref{eq:21}) in p-p collisions for the near- and away side, on the left and right respectively. An obvious splitting effect can be visible between the unlike-sign associated particle pairs, which is the largest for $p/\bar{p}$, and the smallest for $\pi^{+}/\pi^{-}$. In contrast, on the away side the splitting effect is not really present. This indicates that in PYTHIA generated particles does not have connection to the choosen trigger particle in this context. From this picture we can qulitativley say that the strength of the quantum number conservation effects is increasing in the order of charge ($C$), strangeness ($S$) and baryon number ($B$) respectively.

Eq.~(\ref{eq:21}) is plotted for HIJING generated events in the lower panels of Fig.~\ref{fig:PID-triggered_ratio}, as well. A peculiar pattern is present on the near- and away side in the $p_{T,assoc}=2-8$ GeV/c and $p_{T,assoc}=2-10$ GeV/c ranges for charged kaons and protons, where the baryon/meson anomaly was observed at RHIC and at the LHC~\cite{ALICEbaryon-meson}. An opposite trend of the splitting at higher transverse momenta can be attributed to the minijet production embedded in the generator.

\subsection{Identified particle ratios in different trigger $p_T$ bins}
\label{subsec:3}

So far the associated particle spectra were studied in case of wide range of trigger particles' momenta \mbox{$2$ GeV/c $ < p_{T,trig} < 25$ GeV/c}. Here we briefly review the $p_{T,trig}$ dependence of the spectra. For this we plot the differences between the like-sign and unlike-sign particle pairs for the normalized yields $(\mathrm{d}N^{PID}_{assoc}/\mathrm{d}p_{T,assoc})/(\mathrm{d}N^{hadron}_{assoc}/\mathrm{d}p_{T,assoc})$ as plotted in Fig.~\ref{fig:PID-triggered_ratio}. Since positively charged triggers were choosen we can introduce the definition below in order to plot these ratios higher than one:

\footnotesize
\begin{equation}\label{eq:2}
\begin{split}
\left(\frac{\mathrm{d}N^{PID}_{assoc}}{\mathrm{d}p_{T,assoc}}~\Big/~\frac{\mathrm{d}N^{hadron}_{assoc}}{\mathrm{d}p_{T,assoc}}\right)^{(+/-)}~\Big/~\left(\frac{\mathrm{d}N^{PID}_{assoc}}{\mathrm{d}p_{T,assoc}}~\Big/~\frac{\mathrm{d}N^{hadron}_{assoc}}{\mathrm{d}p_{T,assoc}}\right)^{(+/+)} & = \\
= \left(\frac{\mathrm{d}N^{PID}_{assoc}}{\mathrm{d}p_{T,assoc}}\right)^{(+/-)}~\Big/~\left(\frac{\mathrm{d}N^{PID}_{assoc}}{\mathrm{d}p_{T,assoc}}\right)^{(+/+)}&~,
\end{split}
\end{equation}

\begin{figure}[!htb]
		\centering
		\includegraphics[bb = 0 0 531 350, width=\textwidth]{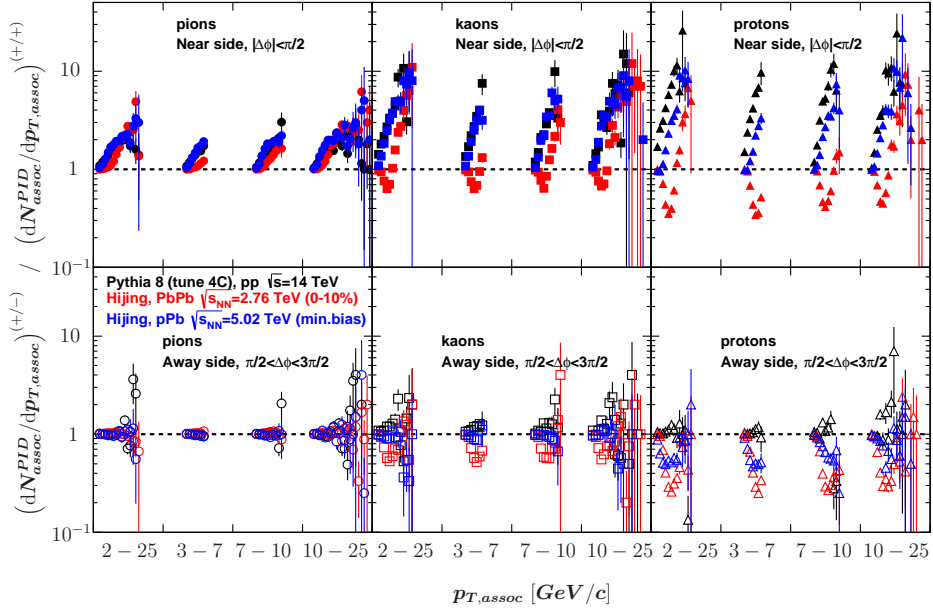}
		\caption{(Color online) Yield differences between the oppositely charged trigger/associated particle pairs. The yield enhancement is relative to the positively charged trigger particles. Different colors refer to different Monte Carlo data sets as indicated on the plot. Upper panel refers to near side, while lower panel refers to away side.}
	\label{fig:PID-triggered_ratio_vs_triggerPtBins}
	\end{figure}

\normalsize
where $(+/-)$ and $(+/+)$ denote the unlike-sign and like-sign trigger/associated particle pairs, respectively. This double ratio is plotted for four different $p_{T,trig}$ trigger bins on the horizontal axis of Fig.~\ref{fig:PID-triggered_ratio_vs_triggerPtBins} in cases of p-p and most central (0-10\%) Pb-Pb collisions. Deviation of the value from $1$ means the relative difference between triggered like- and unlike-sign hadron yields.

In the upper row of Fig.~\ref{fig:PID-triggered_ratio_vs_triggerPtBins} near side ratios are plotted. One can observe how the splits\,--\,in agreement with Fig.~\ref{fig:PID-triggered_ratio}\,--\,evolve as a function of $p_{T,assoc}$ in the near side cone.
In general, the splitting effect is present in all choosen $p_{T,trig}$ bins, $3-7$~GeV/c, $7-10$~GeV/c, and $10-20$~GeV/c, with the same order of magnitude, for the specific partilce species. We can summerize the observed effects as follows:\\

\begin{itemize}
\item In p-p collisions the effect is present for pions, kaons, and protons in the near side cone and the largest for protons with a factor of $10$ effect.
\item In Pb-Pb collisions the effect is present for kaons and protons in the near- and away side cones and the largest for protons.
\item In Pb-Pb collisions the observed trend of the effect for kaons and protons is completely opposite compared to the situatuion in p-p collisions in the intermediate associated transverse momentum spectra 3 GeV/c $ < p_{T,assoc} < $ 7 GeV/c interval.
\end{itemize}

In p-p collisions at the away side the effect is increasing with $p_{T}$ but the magnitude saturates to a constant value. A similar, but larger effect is seen in central Pb-Pb collisions within the $3-10$ GeV/c transverse momentum range, where the Cronin effect~\cite{Cronin:1977} and the baryon/meson anomaly act.

\section{PID-triggered unlike sign-to-like sign associated spectra ratios in p-p and Pb-Pb collisions}
\label{sec:3}

\subsection{Energy dependence in p-p collisions}
\label{subsec:31}

	\begin{figure}[!htb]
		\centering
		\includegraphics[bb= 0 0 477 450, width=0.69\linewidth]{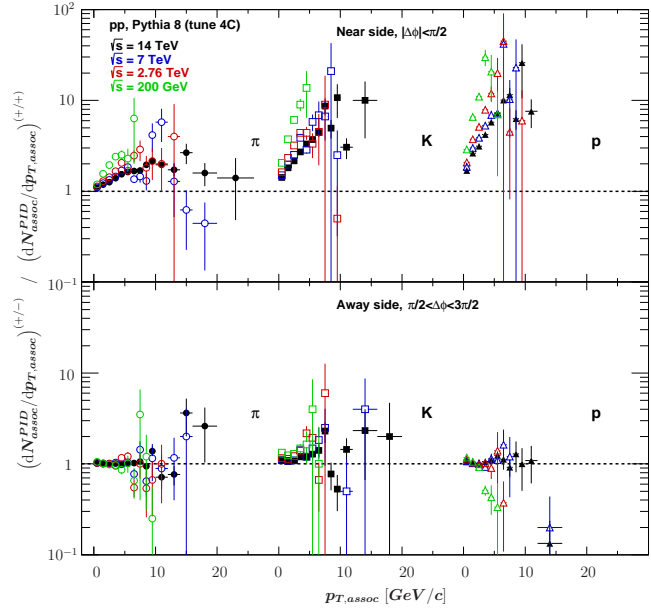}
		\caption{(Color online) Yield differences between the oppositely charged trigger/associated particle pairs for different collision energies.}
	\label{fig:31}
	\end{figure}

In Fig.~\ref{fig:31} we plotted the unlike-to-like sign relative yield ratio defined by Eq.~(\ref{eq:2}) for pions, kaons, and protons with different collision energies $\sqrt{s}$. At the near side (upper panel of the figure) one can see a clear dependence of the splitting on $\sqrt{s}$ as a function of $p_{T,assoc}$ for all particle species. As we go up with $\sqrt{s}$ the effect shows a decreasing trend. This can be related to the fact that the higher the collision energy the more the number of hard interactions. This has effect on the splitting and it is 'washed out'.\\
On the away side we have no such kind of $\sqrt{s}$ dependence, which can be expected based on the argument we had in Sec.~\ref{sec:2}.

\newpage

\subsection{Event multiplicity dependence in p-p collisions}
\label{subsec:32}

	\begin{figure}[!htb]
		\centering
		\includegraphics[bb= 0 0 477 450, width=0.7\linewidth]{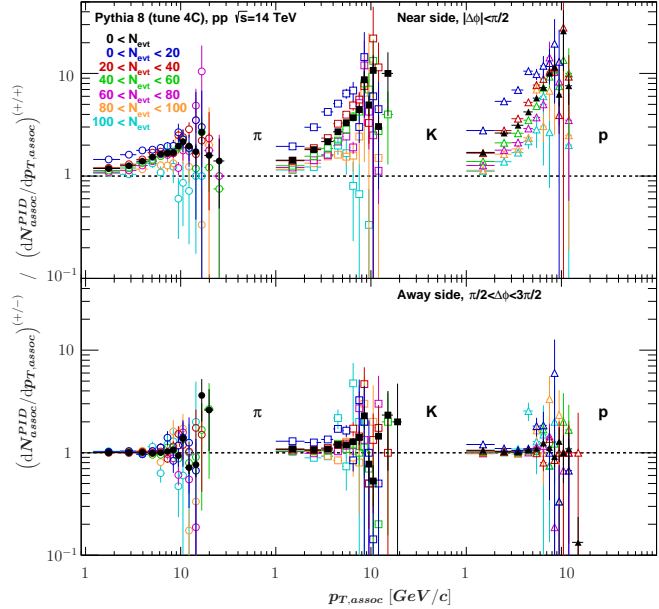}
		\caption{(Color online) Yield differences between the oppositely charged trigger/associated particle pairs for different event multiplicities in $\sqrt{s}=14$ TeV p-p collisions simulated by PYTHIA~8 (tune 4C).}
	\label{fig:32}
	\end{figure}

In Fig.~\ref{fig:32} one can see the event multiplicity dependence of Eq.~(\ref{eq:2}) for pions, kaons, and protons at $\sqrt{s}=14$~TeV in p-p collisions. We observe the same kind of trend in the evolution of the splitting with the multiplicity what have been seen in the previous case, since the number of generated particles depends on the collision energy $\sqrt{s}$.
This case can be treated as a subset of the previous one, becasue here we fix the $\sqrt{s}$ and this way the generated number of particles are bouded. We can conclude that the higher the event multiplicity the smaller the size of the splitting due to increasing underlying event contribution at higher multiplicities.\\
Very similar to the case with $\sqrt{s}$ dependence we state here that the effects on the away side have no obvious dependence with event multiplicity.

\subsection{Centrality dependence in Pb-Pb collisions}
\label{subsec:33}

	\begin{figure}[!htb]
		\centering
		\includegraphics[bb= 0 0 477 450, width=0.7\linewidth]{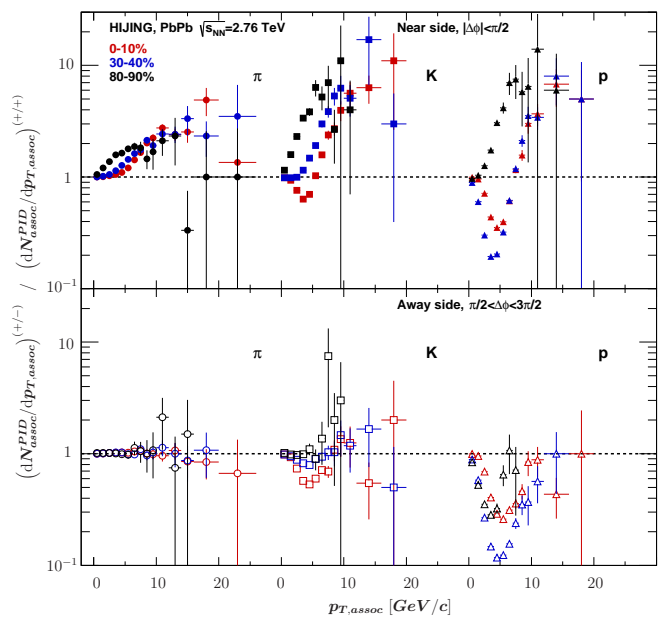}
		\caption{(Color online) Yield differences between the oppositely charged trigger/associated particle pairs for different centralities in Pb-Pb collisions at $\sqrt{s_{NN}}=2.76$~TeV.}
	\label{fig:33}
	\end{figure}

In Fig.~\ref{fig:33} the centrality dependence of Eq.~(\ref{eq:2}) is plotted in Pb-Pb collisions at $\sqrt{s_{NN}}=2.76$~TeV. We expected here a similar effect such as obtained in the previous subsection due to the fact that the multiplicity is related to centrality. In peripheral collisions (black symbols, centrality $80-90\%$) we observe a very similar behaviour for pions as we have seen in Subsec. \ref{subsec:31} for the collision energy dependence in p-p collisions. For kaons and protons there are two major deviations. Firstly, Eq.~(\ref{eq:2}) on the near side both for the kaons and protons starts at the same value as for pions, while in Fig.~\ref{fig:31} at the same collision energy ($\sqrt{s}=2.76$~TeV) we see an obvious inverse deviation from unity in the first $p_{T,assoc}$ bin\,--\,which has the highest value for protons. This deviation turns to be greater than $1$, approaching to highest $p_{T}$ values. Secondly, we do see an inverse evolution too on the away side for kaons and protons.\\
Examining the most central (red symbols, centrality $0-10\%$) and mid central (blue symbols, centrality $30-40\%$) events  we clearly see a reverse trend up to $7-10$~GeV/c, which is attributed to the region of the minijet production. Surprisingly, associated particles on the away side show a stronger effect.

\section{Summary and Outlook}\label{sec:4}

Monte Carlo studies of two-particle azimuthal correlations have been performed
with identified trigger particles and identified associated particles in p-p 
and Pb-Pb collisions at mid-rapidity. 

The presented analysis aims to suggest a possible measurement of the
charge conservation effect via PID-triggered spectra. We think it would
be interesting to see whether measured data behaves similarly as PYTHIA
or HIJING based Monte Carlo generated events. The key issues we have
found in our analysis: 
\begin{itemize}

\item Quantum number conservation can be seen in the near side and we
have got weak signal on the away side, which suggest the correlations
mainly arise from parton cascade and hadronization processes in Monte 
Carlo generated data samples.

\item The effect is getting stronger with multiple quantum number
conservation $\sim 2$ for pions ($C$), $\sim 5$ for kaons ($C+S$), and
$\sim 10$ for proton ($C+B$).

\item The relative PID-triggered yields defined by Eq.~(\ref{eq:2}),
presents increasing splitting up to $p_T<10$ GeV/c and seem
saturating at the highest momenta at the near side, however the lack of
this effect is observed at the away side in PYTHIA simulated p-p
collisions.

\item In case of HIJING-simulated Pb-Pb collisions this splitting effect
is also presents, but inverse trend can be seen in the unlike-to-like
relative PID-triggered yield. This anomalous behavior can be either due
to the minijet production or high-energy nuclear effect (Cronin,
shadowing/anti-shadowing) modeled in the HIJING Monte Carlo generator. 

\item Both collision energy and multiplicity dependence of the effect
were investigated in p-p in addition to the centrality-dependence
simulations in central and peripheral Pb-Pb collisions. Here high-momenta
or low-luminosity events presented larger effect, but this may sensitive
to the background subtraction (here ZYAM) or the mixture of the
soft-hard regime in Monte Carlo event generator cocktails.

\end{itemize}

In summary our aim is to suggest to perform the presented analysis on
real data, where either the nature of the correlations or the goodness
of the Monte Carlo generators can be tested.

\section*{Acknowledgement}

This work was supported by Hungarian OTKA grants, NK106119, K104260, and TET 12 CN-1-2012-0016. Author Gergely G\'abor Barnaf\"oldi also thanks the J\'anos Bolyai Research Scholarship of the Hungarian Academy of Sciences.


\end{document}